\documentclass[%
 reprint,
 amsmath,amssymb,
 aps,
]{revtex4-2}

\DeclareUnicodeCharacter{2212}{-}
\usepackage{amsmath}
\usepackage{caption}
\usepackage{xcolor}
\usepackage{graphicx}
\usepackage{dcolumn}
\usepackage{bm}

\begin{document}

\preprint{APS/123-QED}

\title{Testing $\Lambda$CDM with eBOSS data using a model independent diagnostic}

\author{Arman Shafieloo}
\email{shafieloo@kasi.re.kr}
\author{Sangwoo Park}
\affiliation{Korea Astronomy and Space Science Institute (KASI), 776 Daedeok-daero, Yuseong-gu, Daejeon 34055, Korea}
\affiliation{KASI Campus, University of Science and Technology, 217 Gajeong-ro, Yuseong-gu, Daejeon 34113, Korea}

\author{Varun Sahni}
\affiliation{Inter-University Centre for Astronomy and Astrophysics (IUCAA), Post Bag 4, Ganeshkhind, Pune 411 007, India}

\author{Alexei A. Starobinsky}
\affiliation{L. D. Landau Institute for Theoretical Physics RAS, Chernogolovka, Moscow 142432, Russia}

\date{\today}

\begin{abstract}
The $Om3$ diagnostic \cite{Shafieloo:2012rs} tests the consistency of the cosmological constant as a candidate for dark energy using baryon acoustic oscillation (BAO) data. An important feature of $Om3$ is that it is independent of any parametric assumption for dark energy, neither does it depend upon the dynamics of the Universe during the prerecombination nor postrecombination eras. In other words, $Om3$ can be estimated using BAO observables and used either to confirm or falsify the cosmological constant independently of the value of the Hubble constant $H_0$ (expansion rate at $z=0$), and the comoving sound horizon at the baryon drag epoch, $r_d$ (which is a function of the physics of the Universe prior to recombination). Consequently, $Om3$ can play a key role in identifying the nature of dark energy regardless of the existing tensions in the standard model of cosmology and the possible presence of systematics in some of the data sets. We revisit $Om3$ using the most recent BAO observables from the eBOSS survey in order to test the consistency of the cosmological constant with this data. Our results show the consistency of dark energy being the cosmological constant. Moreover, with eBOSS data, we have achieved a precision of $1.5\%$ for this three-point diagnostic. This demonstrates that $Om3$ can be a potent diagnostic of dark energy when used in conjunction with the high precision data expected from forthcoming large scale structure surveys such as the Dark Energy Spectroscopic Instrument (DESI) and Euclid.     
\end{abstract}

\maketitle

\section{Introduction}

The standard model of cosmology, namely the spatially flat $\Lambda$CDM universe with a power-law form of the primordial scalar perturbation spectrum close to the scale-invariant one, $|n_s(k)-1|\ll 1$, agrees well with most observational data sets including the cosmic microwave background (CMB), distance-redshift relation of type Ia supernovae, data from large scale structure redshift surveys and weak lensing shear data. However, some tension has been noticed while fitting the model to data sets where the posterior of the key cosmological parameters differs substantially. For instance, $\Lambda$CDM predicts a lower value of the Hubble constant when fitting to CMB and baryon acoustic oscillation (BAO) data \cite{Planck:2018vyg, eBOSS:2020yzd} as compared to local measurements of $H_0$ derived from Cepheid-supernova observations \cite{Riess:2016jrr,Riess:2019cxk,Riess:2020fzl}. This is now well-known as the Hubble tension. Since this level of tension has exceeded $5\sigma$ \cite{Riess:2021jrx}, numerous attempts have been made to understand and reconcile it. For instance, attempts have been made to explain the tension by searching for systematics in some of the data sets involved \cite{Freedman:2019jwv,Keeley:2020aym}, such as local measurement of the Hubble constant from Cepheid-supernova observations. (So far, no discrepancies have been reported.) Several theoretical phenomenological models have also been advanced which go beyond the standard model and explain the data using an alternative cosmology \cite{Sahni:2006pa, Keeley:2019esp, Li:2019yem,Ballardini:2020iws}. Some of these models propose changes in the expansion history of the early Universe, while others alter the expansion history of the late Universe. Meanwhile, there have also been attempts to go beyond the power-law form of the primordial fluctuations in order to explain the cosmological tension~\cite{Hazra:2018opk,2020JCAP...09..055K,Hazra:2022rdl,Antony:2022ert}. While some alternative cosmologies do explain the data better than $\Lambda$CDM, none has successfully removed the tension entirely. We believe that in view of the complex nature of the current situation, it is important to consider all possibilities, including the presence of unknown systematics in any of the cosmological data sets, in conjunction with the method of analysis used to test the standard model. 

Given the fluid and somewhat ambiguous situation which currently prevails, it becomes difficult to either rule out or confirm the presence of the exact cosmological constant at a high confidence level. In this context, null diagnostics of $\Lambda$CDM including $Om3$ become relevant.

$Om3$ \cite{Shafieloo:2012rs} is a three-point null diagnostic of concordance cosmology ($\Lambda$CDM). It follows in the footsteps
of its predecessors the two-point null diagnostics $Om$ \cite{Sahni:2008xx,Zunckel:2008ti}
and $Omh^2$ \cite{Sahni:2014ooa}.
$Om3$ is based on the remarkable property  that $Om3 (z_i,z_j,z_k)=1$ only for $\Lambda$CDM (for all redshift combinations of $z_i$). $Om3$  has three additional attractive features. First, it is independent of any parametric or model dependent assumption for dark energy since it directly uses observables from BAO data. Second, the determination of  $Om3$ is independent of the value of the Hubble constant that makes this diagnostic insensitive to the presence of any systematics in measurements of $H_0$. Third, $Om3$ is independent of the physics of the prerecombination Universe and hence is independent of either correct or incorrect assumptions used in the modeling of the early Universe. 

In the following, we explain $Om3$ diagnostic and how to estimate it using BAO observables in Sec.~\ref{sec:Om3}. In Sec.~\ref{sec:results}, we present our results using the eBOSS final cosmology data and in Sec.~\ref{sec:summary}, we conclude.


\section{$Om3$}
\label{sec:Om3}

 The family of $Om$ diagnostics ($Om$, $Om3$, $Omh^2$) was introduced in \cite{Sahni:2008xx,Zunckel:2008ti,Shafieloo:2009hi,Shafieloo:2012rs,Sahni:2014ooa} to distinguish evolving dark energy (DE) from the cosmological constant on the basis of observations of the expansion history; $H(z)$ or $h(z)$. In this paper, we use a specific form of the $Om3$ diagnostic which is tailor made for the analysis of eBOSS data \cite{eBOSS:2020yzd}.
 
The two-point $Om$ diagnostic is defined as follows: 

\begin{equation}
Om(z_j;z_i) = \frac{h^2(z_j)-h^2(z_i)}{(1+z_j)^3 - (1+z_i)^3}\, , ~~h(z) = H(z)/H_0\, ,
\label{eq:om}
\end{equation}
so that \cite{Sahni:2008xx}

\begin{equation}
Om(z;0) \equiv Om(z) = \frac{h^2(z)-1}{(1+z)^3 - 1}~.
\end{equation}

The possibility of using $Om$ as a {\em null diagnostic} follows from the fact that for the cosmological constant 

\begin{equation}
Om(z_j;z_i) = \Omega_{0m}~.
\label{eq:LambdaOm}
\end{equation}

In other words, the value of $Om$ is {\em redshift independent} for the cosmological constant, while for other models of cosmic acceleration $Om(z)$ is redshift dependent.

$Om3$ is a three-point diagnostic which was introduced in \cite{Shafieloo:2012rs} and defined as follows:

\begin{equation}
Om_{\rm ratio}(z_i,z_j,z_i,z_k) := Om3(z_i;z_j;z_k)
= \frac{Om(z_j;z_i)}{Om(z_k;z_i)}~, 
\label{eq:om3}
\end{equation}
where $Om3 = 1$ for $\Lambda$CDM. Keeping the consistency of this equation, we multiply both the numerator and denominator by $H_0^2 r_d^2$ and rewrite $Om3$ as


\begin{multline}
Om3(z_i;z_j;z_k) = \\
\frac{(H(z_j)r_d)^2-(H(z_i)r_d)^2}{(1+z_j)^3-(1+z_i)^3}\bigg/
\frac{(H(z_k)r_d)^2-(H(z_i)r_d)^2}{(1+z_k)^3-(1+z_i)^3}\, , 
\label{eq:om3a}
\end{multline}
where $r_d$ is the comoving sound horizon at the baryon drag epoch,

\begin{equation}
r_d = \int_{z_d}^{\infty} \frac{c_s(z)}{H(z)} dz, 
\label{eq:r_d}
\end{equation}
where $z_d$ is the redshift of the drag epoch, and $c_s$ is the speed of sound. The drag epoch is approximately at $z = 1020$, when the baryons decouple from the photons after recombination. The estimated value of the comoving sound horizon at the baryon drag epoch for the concordance $\Lambda$CDM cosmology fitting Planck CMB data is 147.18 $\pm$ 0.29 Mpc~\cite{Planck:2018vyg}.

Equation (\ref{eq:om3a}) demonstrates that
measurements of $H(z)r_d$ at different redshifts allow one to easily estimate $Om3$ and test the consistency of the data with the spatially flat $\Lambda$CDM model. Performing error propagation and assuming no correlation between different measurements of $H(z)r_d$ one gets
\begin{multline}
\sigma_{Om3(z_i;z_j;z_k)}^2 = \frac{c^2}{(b_k^2-b_i^2)^2} ~~\times
\\ \left[ \left[\frac{2b_i(b_j^2-b_k^2)}{b_k^2-b_i^2} \right]^2\sigma_{b_i}^2+ 4b_j^2\sigma_{b_j}^2+ \left[ \frac{-2b_k(b_j^2-b_i^2)}{b_k^2-b_i^2}\right]^2\sigma_{b_k}^2  \right] 
\label{eq:om3b}
\end{multline}
where $c=\frac{(1+z_k)^3-(1+z_i)^3}{(1+z_j)^3-(1+z_i)^3}$, $b_i=H(z_i)r_d$, $b_j=H(z_j)r_d$, $b_k=H(z_k)r_d$. \\

In our analysis, we shall use five different uncorrelated estimates of $H(z)r_d$ from BOSS and eBOSS observations. 

\section{Results and Analysis}
\label{sec:results}

\begin{table*}
\caption{\label{tab:table1} The BAO + RSD measurements used in this paper \cite{eBOSS:2020yzd}.}
\centering
\begin{tabular}{|c|c|c|c|c|c|}
\hline
& $z_1$ & $z_2$ & $z_3$ & $z_4$ & $z_5$\\
\hline
Source & BOSS galaxy & eBOSS LRG & eBOSS ELG & eBOSS quasar & Ly$\alpha$-Ly$\alpha$ \\
$z_{\text{eff}}$ & 0.38 & 0.70 & 0.85 & 1.48 & 2.33 \\
$D_H(z)/r_d$ & 24.89 $\pm$ 0.58 & 19.78 $\pm$ 0.46 & 19.6 $\pm$ 2.1 & 13.23 $\pm$ 0.47 & 8.93 $\pm$ 0.28 \\
\hline
\end{tabular}
    
\end{table*}

From the final data release of the eBOSS survey, we select five independent and most precise measurements of the quantity $H(z)r_d$ from LRG, ELG, QSO, and Ly$\alpha$. These are the measurements from BOSS and eBOSS using different tracers. In this analysis LRG-Consensus (BAO + Full Shape) (Pk + 2pcf), ELG-full shape Gaussian approximation, QSO combined RSD data consensus, and Ly$\alpha$ auto+cross combined BAO estimations are used~\cite{eBOSS:2020yzd}. 

A summary of the data used in our analysis is given in Table~\ref{tab:table1}. Note that $H(z)r_d=c r_d/D_H(z)$ so that one can trivially estimate $H(z) r_d$ from the data provided by the eBOSS Collaboration. One should also note that we have used measurements that are uncorrelated with respect to each other. $Om3$ can also be used for correlated data but in that case one needs to consider the effect of correlations on error propagation. In other words,  Eq.~(\ref{eq:om3b}) will have more terms reflecting the correlations between the three input measurements. 

On the basis of these five measurements of $H(z)r_d$ from BOSS and eBOSS, one can estimate the $Om3$ diagnostic for ten different combinations of the input measurements. These combinations would be for [$z_1,z_2,z_3$], [$z_1,z_2,z_4$], [$z_1,z_2,z_5$], [$z_1,z_3,z_4$], [$z_1,z_3,z_5$], [$z_1,z_4,z_5$], [$z_2,z_3,z_4$], [$z_2,z_3,z_5$], [$z_2,z_4,z_5$], and [$z_3,z_4,z_5$]. For each of these ten combinations, we then have six different variations of $Om3$ estimation. These variations for the first combination would be $(z_1;z_2;z_3)$, $(z_1;z_3;z_2)$, $(z_2;z_1;z_3)$, $(z_2;z_3;z_1)$, $(z_3;z_1;z_2)$, and $(z_3;z_2;z_1)$. We can see that for the five input measurements, one has 60 different $Om3$ estimations and all of these estimates should equal to unity for $\Lambda$CDM cosmology. One should note that the $Om3$ diagnostic has an asymmetric form, so for a given redshift triplet, the results for $Om3$ will in general depend upon the ordering of the three redshifts. 

Here, 60 variations of $Om3$ can be reduced to 30 variations because some variations are inverse of the others based on the definition of $Om3$ from Eq.~(\ref{eq:om3a})

\begin{equation}
Om3(z_i;z_j;z_k) = \\
\frac{1}{Om3(z_i;z_k;z_j)}.
\label{eq:inverse}
\end{equation}

In addition, the identity with the cyclic permutation of indices follows from the definition of $Om3$ (\ref{eq:om3a}) too,

\begin{equation}
Om3(z_i;z_j;z_k)\cdot Om3(z_j;z_k;z_i)\cdot Om3(z_k;z_i;z_j) \\
= 1.
\label{eq:cyclic}
\end{equation}

By combining Eqs.~(\ref{eq:inverse}) and~(\ref{eq:cyclic}), we get the following relation:

\begin{equation}
Om3(z_k;z_i;z_j) = \frac{Om3(z_j;z_i;z_k)}{Om3(z_i;z_j;z_k)}\, .
\label{eq:cyclic2}
\end{equation}

This produces ten additional algebraic relations for 5 different redshift inputs that leaves only 20  $Om3$ values which cannot be further expressed as products of such quantities and/or their inverses. However, we will present below all 30 variations to show different precision levels. Our results are shown in Table~\ref{fig:table2} with 10 different sub-tables where each sub-table shows the results of the 3 variations of each measurement combination.

In these subtables, the second column presents the estimated value of $Om3$ for the specific redshift combination together with its uncertainties. The third column shows $\sigma$ deviation from the expectation that DE =  $\Lambda$, and the last column presents the precision of the measurement. One can see that the highest deviation from $\Lambda$ = dark energy, that is, the most significant departure from $Om3=1$, arises for the $Om3(z_1; z_2; z_5)$ variation, and that too at the $1.39\sigma$ level.

Overall our results show a good agreement with the hypothesis that dark energy can be the cosmological constant and the number of the estimates with deviation larger than $1\sigma$ from the expectations of the $\Lambda$, which are six variations for $Om3(z_1,z_2,z_4)$ and $Om3(z_1,z_2,z_5)$ among the total 30 variations also seems reasonable. However, we should note here that the maximum deviation from the premise that the cosmological constant is dark energy is for the combination with the highest precision. This $1.39\sigma$ deviation may be taken as a mild hint of inconsistency with $\Lambda$CDM and its potential significance can only be checked by further observations and analysis.

Our results also demonstrate that one can achieve an impressive precision of about $1.5\%$ in the estimation of $Om3$ for the $Om3(z_5; z_1; z_2)$ variation. This is due to the relatively high precision of BAO data at these redshifts and the relatively larger spacing between two consecutive redshift bins, both of which significantly improve the efficiency of the $Om3$ diagnostic. We expect this precision to be improved further with the forthcoming data from the Dark Energy Spectroscopic Instrument (DESI)~\cite{DESI:2016fyo} and Euclid~\cite{Euclid:2019clj}.

\begin{figure*}
\parbox[t]{7cm}{\null
\centering
  \captionof{table}[t]{Results for $\sqrt{Om(z_i;z_j)} H_0 r_d$ for different redshift combinations. For the concordance model, this quantity is equivalent to $\sqrt{\Omega_{0m}} H_0 r_d$.}
  \vskip\abovecaptionskip
\begin{tabular}{|c|c|}
\hline

& $\sqrt{Om(z_i;z_j)} H_0 r_d$ (km/s) \\

\hline
$(z_1; z_2)$& 6086.28 $\pm$ 454.60 \\
$(z_1; z_3)$& 4898.80 $\pm$ 1394.11 \\
$(z_1; z_4)$& 5401.91 $\pm$ 272.03 \\
$(z_1; z_5)$& 5350.73 $\pm$ 193.44 \\
$(z_2; z_3)$& 1728.54 $\pm$ 10451.82 \\
$(z_2; z_4)$& 5238.63 $\pm$ 350.91 \\
$(z_2; z_5)$& 5294.33 $\pm$ 210.87 \\
$(z_3; z_4)$& 5597.50 $\pm$ 620.80 \\
$(z_3; z_5)$& 5402.88 $\pm$ 262.11 \\
$(z_4; z_5)$& 5320.70 $\pm$ 344.87 \\
\hline

\end{tabular}
\label{fig:table3}
}
\parbox[t]{10cm}{\null
  \centering
\includegraphics[width = 1.0\linewidth]{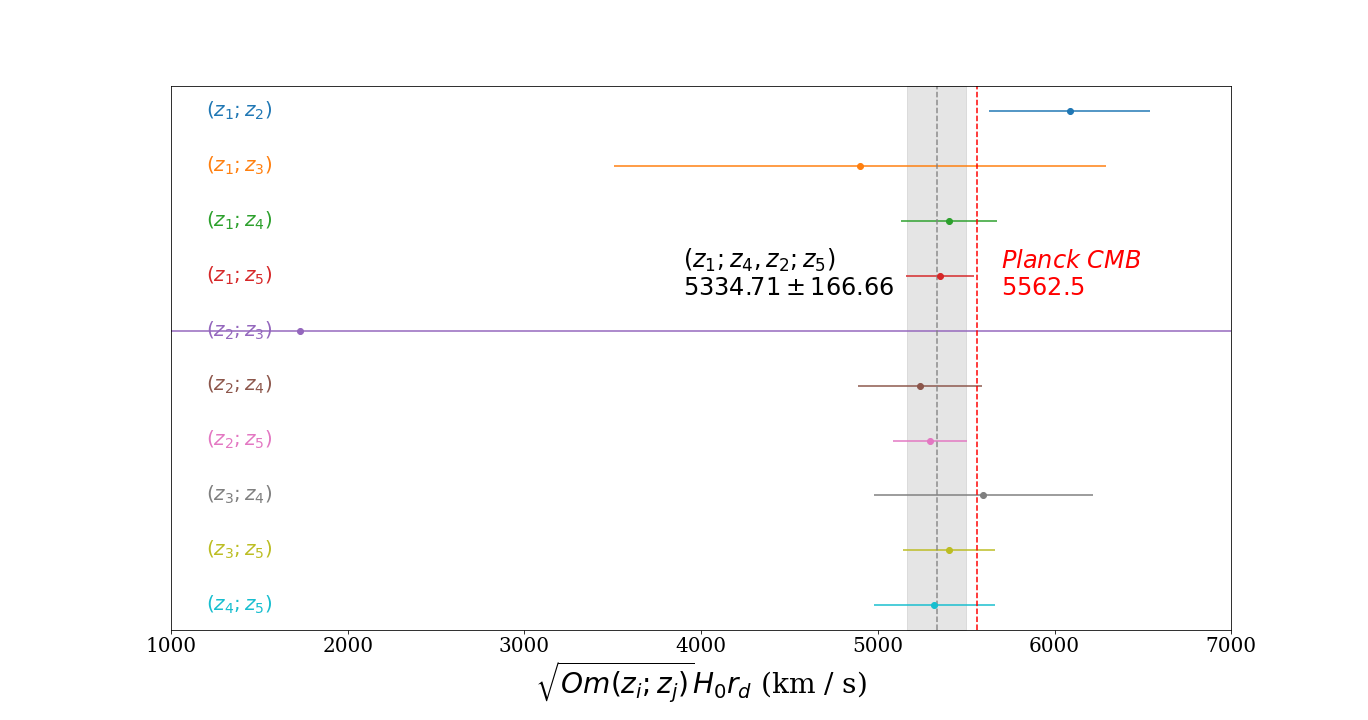}
\captionof{figure}[t]{Results presented in Table~\ref{fig:table3} for $\sqrt{Om(z_i;z_j)} H_0 r_d$ are shown for different redshift combinations. For the case of the concordance model, this quantity is equivalent to $\sqrt{\Omega_{0m}} H_0 r_d$. The gray vertical band shows the result for the ($z_1;z_4$) and ($z_2;z_5$) combinations (the two independent estimates having the highest precision). The red vertical line shows $\sqrt{\Omega_{0m}} H_0 r_d=5562.5$ km/s for the best-fit concordance model to the Planck CMB data~\cite{Planck:2018vyg}. }  
  
\label{fig:fig1}}
\end{figure*}

There is also one interesting byproduct of our analysis which rests in the physical meaning of the numerator (or denominator) of the equation (\ref{eq:om3a}). The square root of the numerator (or denominator) of~(\ref{eq:om3a}) is equivalent to $\sqrt{Om(z_i;z_j)} H_0 r_d$, which, for the spatially flat $\Lambda$CDM model reduces to $\sqrt{\Omega_{0m}} H_0 r_d$. Here ${Om(z_i;z_j)}$ is the $Om$ diagnostic defined in~(\ref{eq:om}). In other words, the BAO observables can provide us with a direct estimate of a quantity that connects some key cosmological parameters without any model assumption. Results for $\sqrt{Om(z_i;z_j)} H_0 r_d$ are presented in Table~\ref{fig:table3} and Fig.~\ref{fig:fig1}. We can see that the results for all different redshift combinations agree with each other. The grey band represents the combined estimate of the quantity using two independent redshift combinations of ($z_1;z_4$) and ($z_2;z_5$) where we can get $\sqrt{Om(z_i;z_j)} H_0 r_d=5334.71 \pm 166.66 $ km/s. As mentioned above, for the case of the concordance model, this should be equivalent to $\sqrt{\Omega_{0m}} H_0 r_d$ and this is in good agreement with the best-fit estimate of the concordance model fit to Planck CMB data where we get $\sqrt{\Omega_{0m}} H_0 r_d = 5562.5 $ km/s ($1.4\sigma$ consistency)~\cite{Planck:2018vyg}. This comparison can be interpreted as a completely independent consistency test of the BAO and CMB data (temperature and polarization combined) within the context of the standard model.

\begin{table*}
 \caption{Results for $Om3$ diagnostics for different combinations and for all variations. Redshifts are given in Table~\ref{tab:table1}.}
\centering
\begin{tabular}{ccccllcccc}
\cline{1-4} \cline{7-10}
\multicolumn{1}{|c|}{$Om3(z_1,z_2,z_3)$} & \multicolumn{1}{c|}{$Om3$}                & \multicolumn{1}{c|}{$\dfrac{Om3 - 1}{\sigma}$} & \multicolumn{1}{c|}{$\dfrac{\sigma}{Om3}$} &  & \multicolumn{1}{l|}{} & \multicolumn{1}{c|}{$Om3(z_1,z_4,z_5)$} & \multicolumn{1}{c|}{$Om3$}                & \multicolumn{1}{c|}{$\dfrac{Om3 - 1}{\sigma}$} & \multicolumn{1}{c|}{$\dfrac{\sigma}{Om3}$} \\ \cline{1-4} \cline{7-10} 
\multicolumn{1}{|c|}{$Om3(z_1;z_2;z_3)$} & \multicolumn{1}{c|}{1.544 $\pm$ 0.892}    & \multicolumn{1}{c|}{0.609}                     & \multicolumn{1}{c|}{0.578}                 &  & \multicolumn{1}{l|}{} & \multicolumn{1}{c|}{$Om3(z_1;z_4;z_5)$} & \multicolumn{1}{c|}{1.019 $\pm$ 0.125}    & \multicolumn{1}{c|}{0.153}                     & \multicolumn{1}{c|}{0.123}                 \\
\multicolumn{1}{|c|}{$Om3(z_2;z_1;z_3)$} & \multicolumn{1}{c|}{12.398 $\pm$ 150.268} & \multicolumn{1}{c|}{0.076}                     & \multicolumn{1}{c|}{12.120}                &  & \multicolumn{1}{l|}{} & \multicolumn{1}{c|}{$Om3(z_4;z_1;z_5)$} & \multicolumn{1}{c|}{1.031 $\pm$ 0.203}    & \multicolumn{1}{c|}{0.152}                     & \multicolumn{1}{c|}{0.197}                 \\
\multicolumn{1}{|c|}{$Om3(z_3;z_1;z_2)$} & \multicolumn{1}{c|}{8.032 $\pm$ 92.708}   & \multicolumn{1}{c|}{0.076}                     & \multicolumn{1}{c|}{11.542}                &  & \multicolumn{1}{l|}{} & \multicolumn{1}{c|}{$Om3(z_5;z_1;z_4)$} & \multicolumn{1}{c|}{1.011 $\pm$ 0.075}    & \multicolumn{1}{c|}{0.152}                     & \multicolumn{1}{c|}{0.074}                 \\ \cline{1-4} \cline{7-10} 
                                         &                                           &                                                &                                            &  &                       &                                         &                                           &                                                &                                            \\
                                         &                                           &                                                &                                            &  &                       &                                         &                                           &                                                &                                            \\ \cline{1-4} \cline{7-10} 
\multicolumn{1}{|c|}{$Om3(z_1,z_2,z_4)$} & \multicolumn{1}{c|}{$Om3$}                & \multicolumn{1}{c|}{$\dfrac{Om3 - 1}{\sigma}$} & \multicolumn{1}{c|}{$\dfrac{\sigma}{Om3}$} &  & \multicolumn{1}{l|}{} & \multicolumn{1}{c|}{$Om3(z_2,z_3,z_4)$} & \multicolumn{1}{c|}{$Om3$}                & \multicolumn{1}{c|}{$\dfrac{Om3 - 1}{\sigma}$} & \multicolumn{1}{c|}{$\dfrac{\sigma}{Om3}$} \\ \cline{1-4} \cline{7-10} 
\multicolumn{1}{|c|}{$Om3(z_1;z_2;z_4)$} & \multicolumn{1}{c|}{1.269 $\pm$ 0.218}    & \multicolumn{1}{c|}{1.235}                     & \multicolumn{1}{c|}{0.172}                 &  & \multicolumn{1}{l|}{} & \multicolumn{1}{c|}{$Om3(z_2;z_3;z_4)$} & \multicolumn{1}{c|}{0.109 $\pm$ 1.316}    & \multicolumn{1}{c|}{$-$0.677}                  & \multicolumn{1}{c|}{12.086}                \\
\multicolumn{1}{|c|}{$Om3(z_2;z_1;z_4)$} & \multicolumn{1}{c|}{1.350 $\pm$ 0.301}    & \multicolumn{1}{c|}{1.162}                     & \multicolumn{1}{c|}{0.223}                 &  & \multicolumn{1}{l|}{} & \multicolumn{1}{c|}{$Om3(z_3;z_2;z_4)$} & \multicolumn{1}{c|}{0.095 $\pm$ 1.170}    & \multicolumn{1}{c|}{$-$0.773}                  & \multicolumn{1}{c|}{12.269}                \\
\multicolumn{1}{|c|}{$Om3(z_4;z_1;z_2)$} & \multicolumn{1}{c|}{1.063 $\pm$ 0.054}    & \multicolumn{1}{c|}{1.162}                     & \multicolumn{1}{c|}{0.051}                 &  & \multicolumn{1}{l|}{} & \multicolumn{1}{c|}{$Om3(z_4;z_2;z_3)$} & \multicolumn{1}{c|}{0.876 $\pm$ 0.161}    & \multicolumn{1}{c|}{$-$0.773}                  & \multicolumn{1}{c|}{0.183}                 \\ \cline{1-4} \cline{7-10} 
                                         &                                           &                                                &                                            &  &                       &                                         &                                           &                                                &                                            \\
                                         &                                           &                                                &                                            &  &                       &                                         &                                           &                                                &                                            \\ \cline{1-4} \cline{7-10} 
\multicolumn{1}{|c|}{$Om3(z_1,z_2,z_5)$} & \multicolumn{1}{c|}{$Om3$}                & \multicolumn{1}{c|}{$\dfrac{Om3 - 1}{\sigma}$} & \multicolumn{1}{c|}{$\dfrac{\sigma}{Om3}$} &  & \multicolumn{1}{l|}{} & \multicolumn{1}{c|}{$Om3(z_2,z_3,z_5)$} & \multicolumn{1}{c|}{$Om3$}                & \multicolumn{1}{c|}{$\dfrac{Om3 - 1}{\sigma}$} & \multicolumn{1}{c|}{$\dfrac{\sigma}{Om3}$} \\ \cline{1-4} \cline{7-10} 
\multicolumn{1}{|c|}{$Om3(z_1;z_2;z_5)$} & \multicolumn{1}{c|}{1.294 $\pm$ 0.210}    & \multicolumn{1}{c|}{1.397}                     & \multicolumn{1}{c|}{0.163}                 &  & \multicolumn{1}{l|}{} & \multicolumn{1}{c|}{$Om3(z_2;z_3;z_5)$} & \multicolumn{1}{c|}{0.107 $\pm$ 1.289}    & \multicolumn{1}{c|}{$-$0.693}                  & \multicolumn{1}{c|}{12.091}                \\
\multicolumn{1}{|c|}{$Om3(z_2;z_1;z_5)$} & \multicolumn{1}{c|}{1.322 $\pm$ 0.235}    & \multicolumn{1}{c|}{1.367}                     & \multicolumn{1}{c|}{0.178}                 &  & \multicolumn{1}{l|}{} & \multicolumn{1}{c|}{$Om3(z_3;z_2;z_5)$} & \multicolumn{1}{c|}{0.102 $\pm$ 1.243}    & \multicolumn{1}{c|}{$-$0.722}                  & \multicolumn{1}{c|}{12.148}                \\
\multicolumn{1}{|c|}{$Om3(z_5;z_1;z_2)$} & \multicolumn{1}{c|}{1.021 $\pm$ 0.016}    & \multicolumn{1}{c|}{1.367}                     & \multicolumn{1}{c|}{0.015}                 &  & \multicolumn{1}{l|}{} & \multicolumn{1}{c|}{$Om3(z_5;z_2;z_3)$} & \multicolumn{1}{c|}{0.960 $\pm$ 0.055}    & \multicolumn{1}{c|}{$-$0.722}                  & \multicolumn{1}{c|}{0.057}                 \\ \cline{1-4} \cline{7-10} 
                                         &                                           &                                                &                                            &  &                       &                                         &                                           &                                                &                                            \\
                                         &                                           &                                                &                                            &  &                       &                                         &                                           &                                                &                                            \\ \cline{1-4} \cline{7-10} 
\multicolumn{1}{|c|}{$Om3(z_1,z_3,z_4)$} & \multicolumn{1}{c|}{$Om3$}                & \multicolumn{1}{c|}{$\dfrac{Om3 - 1}{\sigma}$} & \multicolumn{1}{c|}{$\dfrac{\sigma}{Om3}$} &  & \multicolumn{1}{l|}{} & \multicolumn{1}{c|}{$Om3(z_2,z_4,z_5)$} & \multicolumn{1}{c|}{$Om3$}                & \multicolumn{1}{c|}{$\dfrac{Om3 - 1}{\sigma}$} & \multicolumn{1}{c|}{$\dfrac{\sigma}{Om3}$} \\ \cline{1-4} \cline{7-10} 
\multicolumn{1}{|c|}{$Om3(z_1;z_3;z_4)$} & \multicolumn{1}{c|}{0.822 $\pm$ 0.473}    & \multicolumn{1}{c|}{$-$0.375}                  & \multicolumn{1}{c|}{0.576}                 &  & \multicolumn{1}{l|}{} & \multicolumn{1}{c|}{$Om3(z_2;z_4;z_5)$} & \multicolumn{1}{c|}{0.979 $\pm$ 0.150}    & \multicolumn{1}{c|}{$-$0.140}                  & \multicolumn{1}{c|}{0.153}                 \\
\multicolumn{1}{|c|}{$Om3(z_3;z_1;z_4)$} & \multicolumn{1}{c|}{0.766 $\pm$ 0.581}    & \multicolumn{1}{c|}{$-$0.403}                  & \multicolumn{1}{c|}{0.759}                 &  & \multicolumn{1}{l|}{} & \multicolumn{1}{c|}{$Om3(z_4;z_2;z_5)$} & \multicolumn{1}{c|}{0.969 $\pm$ 0.217}    & \multicolumn{1}{c|}{$-$0.141}                  & \multicolumn{1}{c|}{0.224}                 \\
\multicolumn{1}{|c|}{$Om3(z_4;z_1;z_3)$} & \multicolumn{1}{c|}{0.931 $\pm$ 0.170}    & \multicolumn{1}{c|}{$-$0.403}                  & \multicolumn{1}{c|}{0.183}                 &  & \multicolumn{1}{l|}{} & \multicolumn{1}{c|}{$Om3(z_5;z_2;z_4)$} & \multicolumn{1}{c|}{0.990 $\pm$ 0.070}    & \multicolumn{1}{c|}{$-$0.141}                  & \multicolumn{1}{c|}{0.071}                 \\ \cline{1-4} \cline{7-10} 
                                         &                                           &                                                &                                            &  &                       &                                         &                                           &                                                &                                            \\
                                         &                                           &                                                &                                            &  &                       &                                         &                                           &                                                &                                            \\ \cline{1-4} \cline{7-10} 
\multicolumn{1}{|c|}{$Om3(z_1,z_3,z_5)$} & \multicolumn{1}{c|}{$Om3$}                & \multicolumn{1}{c|}{$\dfrac{Om3 - 1}{\sigma}$} & \multicolumn{1}{c|}{$\dfrac{\sigma}{Om3}$} &  & \multicolumn{1}{l|}{} & \multicolumn{1}{c|}{$Om3(z_3,z_4,z_5)$} & \multicolumn{1}{c|}{$Om3$}                & \multicolumn{1}{c|}{$\dfrac{Om3 - 1}{\sigma}$} & \multicolumn{1}{c|}{$\dfrac{\sigma}{Om3}$} \\ \cline{1-4} \cline{7-10} 
\multicolumn{1}{|c|}{$Om3(z_1;z_3;z_5)$} & \multicolumn{1}{c|}{0.838 $\pm$ 0.480}     & \multicolumn{1}{c|}{$-$0.337}                  & \multicolumn{1}{c|}{0.573}                 &  & \multicolumn{1}{l|}{} & \multicolumn{1}{c|}{$Om3(z_3;z_4;z_5)$} & \multicolumn{1}{c|}{1.073 $\pm$ 0.211}    & \multicolumn{1}{c|}{0.348}                     & \multicolumn{1}{c|}{0.196}                 \\
\multicolumn{1}{|c|}{$Om3(z_3;z_1;z_5)$} & \multicolumn{1}{c|}{0.822 $\pm$ 0.518}    & \multicolumn{1}{c|}{$-$0.344}                  & \multicolumn{1}{c|}{0.630}                 &  & \multicolumn{1}{l|}{} & \multicolumn{1}{c|}{$Om3(z_4;z_3;z_5)$} & \multicolumn{1}{c|}{1.107 $\pm$ 0.316}    & \multicolumn{1}{c|}{0.338}                     & \multicolumn{1}{c|}{0.286}                 \\
\multicolumn{1}{|c|}{$Om3(z_5;z_1;z_3)$} & \multicolumn{1}{c|}{0.981 $\pm$ 0.056}    & \multicolumn{1}{c|}{$-$0.344}                  & \multicolumn{1}{c|}{0.057}                 &  & \multicolumn{1}{l|}{} & \multicolumn{1}{c|}{$Om3(z_5;z_3;z_4)$} & \multicolumn{1}{c|}{1.031 $\pm$ 0.092}    & \multicolumn{1}{c|}{0.338}                     & \multicolumn{1}{c|}{0.089}                 \\ \cline{1-4} \cline{7-10} 
\end{tabular}

    \label{fig:table2}
\end{table*}

\section{Summary}
\label{sec:summary}

The $Om3$ diagnostic probes departures of dark energy from concordance $\Lambda$CDM using direct observables of BAO surveys. It is independent of any parametric form for dark energy or expansion history and can be estimated independently of the value of the Hubble constant and the baryon drag epoch, $r_d$. Hence $Om3$ is immune to possible systematics in the measurements of these important observational quantities associated with the local universe and the prerecombination era, respectively. 

Our results demonstrate a good consistency of the eBOSS final cosmology data release with the cosmological constant as dark energy. Most of the variations of $Om3(z_i;z_j;z_k)$ show consistency with the $\Lambda$CDM prediction $Om3=1$. The most significant departure from $Om3=1$, amongst all of the 30 variations of the redshift triplet $(z_i;z_j;z_k)$ (from which 20 cannot be further expressed as products of these quantities and/or their inverses) is at the $1.39\,\sigma$ level. The precision of the $Om3(z_i;z_j;z_k)$ variations can vary significantly from one combination of the triplet $(z_i;z_j;z_k)$ to another. In general, if there is a large spacing between the three input data points, $z_i,z_j,z_k$, and the data points have high quality measurements (high precision), we expect to derive $Om3$ with high precision. On the other hand, if the data points (at least two of them) are close to each other in redshift space or/and BAO determinations at these data points have large uncertainties, we expect $Om3$ to have poor precision. While many of the variations of the redshift triplet $(z_i;z_j;z_k)$ led to large uncertainties in the $Om3$ diagnostic, the maximum precision we achieved for $Om3$ using eBOSS data is in fact very promising and impressive. For $Om3(z_5;z_1;z_2)$, the BAO data at $z_1=0.38$, $z_2=0.70$, and $z_5=2.33$ is known to high precision, additionally, there is a large spacing between adjacent redshifts. These factors result in a $1.5\%$ precision determination of $Om3$ for this combination of the data. While the current precision is not yet very high, this can be potentially a significant result since it demonstrates that $Om3$ might be able to differentiate evolving dark energy models from $\Lambda$CDM with the help of high precision data from forthcoming galaxy redshift surveys such as DESI and Euclid. 

We should emphasize that $Om3$ is sensitive to the change in the shape of the expansion history, and two models with comparable $\chi^2$ likelihoods to the BAO data may show different consistencies to the $Om3$ estimations. This is very distinct from using any one-point diagnostics to evaluate deviations from the predictions of a given model, e.g. the spatially flat $\Lambda$CDM one. In Appendix A, we discuss this point in some detail. Also looking if a given value of $Om3$ for some three redshift points is equal to unity or not (taking error bars into account, of course), we check only that these three values of $H(z)$ can (or, cannot) belong to $\Lambda$CDM with some parameters $H_0$ and $\Omega_m$, but we do not require that these parameters should be the {\em same} (inside error bars) for $Om3$ values estimated at all other choices of three redshifts. So, $Om3$ is more sensitive to deviations from $\Lambda$CDM occurring near some special points which can be rather far from each other.

This would guide us to another potential possibility in order to use $Om3$ for model selection and parameter estimation. While $Om3$ is equal to unity for the case of cosmological constant with $w=-1$ (for all matter densities), for any given dark energy model with $w(z)\ne-1$, the value of $Om3$ would be a varying quantity depending on the input redshifts and the assumed cosmology [$w(z)$ and $\Omega_{0m}$]. These theoretical expectations of $Om3$ for any given model and any redshift combination can be calculated and compared with the estimates of $Om3$ from observations. It is beyond the scope of this work to use $Om3$ for model selection and parameter estimation, but we can note here that there is an interesting potential in this diagnostic to set additional limits on different dark energy models.     

An additional interesting point in this analysis is that for $\Lambda$CDM the numerator / denominator of the equation associated with $Om3$, namely (\ref{eq:om3a}), reduce to ${Om(z_i;z_j)} H_0^2 r_d^2$, with ${Om(z_i;z_j)}$
defined in (\ref{eq:om}). 

Hence, having five independent measurements of $H(z)r_d$ leads to ten different measurements of this quantity or subsequently $\sqrt{Om(z_i;z_j)} H_0 r_d$. Interestingly this quantity is equivalent to $\sqrt{\Omega_{0m}} H_0 r_d$ for the spatially flat $\Lambda$CDM model, and this allows us to test the consistency of different data, such as BAO and CMB within the context of the concordance model.

\section*{Acknowledgments}

This work was supported by the high performance computing cluster Seondeok at the Korea Astronomy and Space Science Institute (KASI). A.~S. would like to acknowledge the support by National Research Foundation of Korea NRF-2021M3F7A1082053 and the support of the Korea Institute for Advanced Study (KIAS) grant funded by the government of Korea. V.~S. was partially supported by the J. C. Bose Fellowship of Department of Science and Technology, Government of India. A.~A.~S. was partly supported by the Project No. 0033-2019-0005 of the Russian Ministry of Science and Higher Education.

\begin{table*}
\caption{\label{table:tableapp} The results for one-point functions $f(z)$ and the three-point function $Om3$ are shown for two models of dark energy namely $w$CDM (\textit{top}) and $\Lambda$CDM (\textit{bottom}) with arbitrary choices of $H_0$ and $A$. One can see the sensitivity of the $f(z)$ values to the choices of $H_0$ and $A$ while $Om3$ is independent of these assumptions.}

\centering

  
1. $w$CDM ($w = -0.905$, $\Omega_m = 0.285$)
\vspace{10pt}

$H_0$ = 67.37, $A$ = 0.3

\begin{tabular}{|c|c|c|c|c|c|c|}
\hline

\rule{0pt}{10pt}$f(z_1)$ & $f(z_2)$ & $f(z_3)$ & $f(z_4)$ & $f(z_5)$ &  $\overline{f(z)}$ & $f(z)_\mathrm{RMS}$\\ \hline

 0.044       &    0.058      &     0.057     &     $-0.003$     &     $-0.246$  & $-0.018$ & 0.118 \\ \hline 
\end{tabular}
\vspace{10pt}

$H_0$ = 67.37, $A$ = 0.285

\begin{tabular}{|c|c|c|c|c|c|c|}
\hline
\rule{0pt}{10pt}$f(z_1)$ & $f(z_2)$ & $f(z_3)$ & $f(z_4)$ & $f(z_5)$ &  $\overline{f(z)}$ & $f(z)_\mathrm{RMS}$\\ \hline
  0.068       &    0.117      &     0.137     &     0.211     &     0.292  & 0.165 & 0.183   \\ \hline
\end{tabular}
\vspace{10pt}

$H_0$ = 73.2, $A$ = 0.3

\begin{tabular}{|c|c|c|c|c|c|c|}
\hline
\rule{0pt}{10pt}$f(z_1)$ & $f(z_2)$ & $f(z_3)$ & $f(z_4)$ & $f(z_5)$ &  $\overline{f(z)}$ & $f(z)_\mathrm{RMS}$\\ \hline
  $-0.190$       &    $-0.283$      &     $-0.349$     &     $-0.809$     &     $-2.010$ & $-0.728$ & 0.993    \\ \hline
\end{tabular}
\vspace{10pt}

$H_0$ = 73.2, $A$ = 0.285

\begin{tabular}{|c|c|c|c|c|c|c|}
\hline
\rule{0pt}{10pt}$f(z_1)$ & $f(z_2)$ & $f(z_3)$ & $f(z_4)$ & $f(z_5)$ &  $\overline{f(z)}$ & $f(z)_\mathrm{RMS}$\\ \hline
  $-0.165$       &    $-0.224$      &     $-0.269$     &     $-0.595$     &     $-1.471$ & $-0.545$ & 0.730   \\ \hline
\end{tabular}
\vspace{10pt}

\begin{tabular}{|c|c|}
\hline
  $Om3$      &  Equation (\ref{eq:(4)})  \\ \hline
$Om3(z_1;z_2;z_3)$    & 1.008  \\ \hline
$Om3(z_1;z_2;z_4)$    & 1.033  \\ \hline
$Om3(z_1;z_2;z_5)$    & 1.050  \\ \hline
$Om3(z_1;z_3;z_4)$    & 1.024  \\ \hline
$Om3(z_1;z_3;z_5)$    & 1.041 \\ \hline
$Om3(z_1;z_4;z_5)$    & 1.016  \\ \hline
$Om3(z_2;z_3;z_4)$    & 1.018 \\ \hline
$Om3(z_2;z_3;z_5)$    & 1.030  \\ \hline
$Om3(z_2;z_4;z_5)$    & 1.013  \\ \hline
$Om3(z_3;z_4;z_5)$    & 1.011  \\ \hline

\end{tabular}

\vspace{20pt}

  \centering
2. $\Lambda$CDM ($w = -1.0$, $\Omega_m = 0.3$)
\vspace{10pt}

$H_0$ = 67.37, $A$ = 0.3

\begin{tabular}{|c|c|c|c|c|c|c|}
\hline
\rule{0pt}{10pt}$f(z_1)$ & $f(z_2)$ & $f(z_3)$ & $f(z_4)$ & $f(z_5)$ &  $\overline{f(z)}$ & $f(z)_\mathrm{RMS}$\\ \hline
  0.000       &    0.000      &     0.000     &     0.000     &     0.000 & 0.000 & 0.000    \\ \hline
\end{tabular}
\vspace{10pt}

$H_0$ = 67.37, $A$ = 0.285

\begin{tabular}{|c|c|c|c|c|c|c|}
\hline
\rule{0pt}{10pt}$f(z_1)$ & $f(z_2)$ & $f(z_3)$ & $f(z_4)$ & $f(z_5)$ &  $\overline{f(z)}$ & $f(z)_\mathrm{RMS}$\\ \hline
  0.024       &    0.059    &     0.080     &     0.214     &     0.539   & 0.183 & 0.263  \\ \hline
\end{tabular}
\vspace{10pt}

$H_0$ = 73.2, $A$ = 0.3

\begin{tabular}{|c|c|c|c|c|c|c|}
\hline
\rule{0pt}{10pt}$f(z_1)$ & $f(z_2)$ & $f(z_3)$ & $f(z_4)$ & $f(z_5)$ &  $\overline{f(z)}$ & $f(z)_\mathrm{RMS}$\\ \hline
  $-0.228$       &    $-0.332$      &     $-0.398$     &     $-0.807$     &     $-1.801$  & $-0.713$ & 0.918   \\ \hline
\end{tabular}
\vspace{10pt}

$H_0$ = 73.2, $A$ = 0.285

\begin{tabular}{|c|c|c|c|c|c|c|}
\hline
\rule{0pt}{10pt}$f(z_1)$ & $f(z_2)$ & $f(z_3)$ & $f(z_4)$ & $f(z_5)$ &  $\overline{f(z)}$ & $f(z)_\mathrm{RMS}$\\ \hline
  $-0.203$       &    $-0.274$      &     $-0.318$     &     $-0.593$     &     $-1.262$  & $-0.530$ & 0.658 \\ \hline
\end{tabular}
\vspace{10pt}

\begin{tabular}{|c|c|c|}
\hline
  $Om3$      &   Equation (\ref{eq:(4)})  \\ \hline
$Om3(z_1;z_2;z_3)$    & 1.000  \\ \hline
$Om3(z_1;z_2;z_4)$    & 1.000  \\ \hline
$Om3(z_1;z_2;z_5)$    & 1.000  \\ \hline
$Om3(z_1;z_3;z_4)$    & 1.000  \\ \hline
$Om3(z_1;z_3;z_5)$    & 1.000 \\ \hline
$Om3(z_1;z_4;z_5)$    & 1.000  \\ \hline
$Om3(z_2;z_3;z_4)$    & 1.000 \\ \hline
$Om3(z_2;z_3;z_5)$    & 1.000  \\ \hline
$Om3(z_2;z_4;z_5)$    & 1.000  \\ \hline
$Om3(z_3;z_4;z_5)$    & 1.000  \\ \hline

\end{tabular}
  
\end{table*}

\appendix
\section{$Om3$ IN COMPARISON WITH POSSIBLE ONE-POINT DIAGNOSTICS}

As we emphasized, $Om3$ is a {\em model independent} diagnostic of cosmological constant as dark energy, and its value derived directly from the BAO observables, $Om3\equiv 1$ is independent of cosmological parameters. One may argue that having a large number of observables, we might be able to evaluate the deviation from the expectations of the $\Lambda$ dark energy in a more direct way by introducing a deviation function at each data redshift. To make this comparison, let us consider a discrete function $f(z)$ describing the deviation of the expansion history from the $\Lambda$CDM model at each data redshift. This would practically evaluate the deviation from the expectations of the spatially flat $\Lambda$CDM model with a set of one-point functions. We would have

\begin{equation}
\frac{H^2(z)}{H_0^2} = 1 - A + A(1+z)^3 + f(z),
\label{eq:(1)}
\end{equation}
\\
note that in the case of $f(z)=0$ we get the standard $\Lambda$CDM model where $A$ would represent $\Omega_m$. Therefore, 
\begin{equation}
f(z) = \frac{H^2(z)}{H_0^2} - 1 - A \{(1+z)^3 - 1 \},
\label{eq:(2)}
\end{equation}
where no assumptions about $f(z)$ are made, and we can take $H_0$ and $A$ as free parameters. If we consider five different redshifts from $z_1$ to $z_5$ just to make it similar to what we did in this analysis, then we would have two free parameters and five values of $f(z)$ from BAO observables. Thus, there are seven correlated parameters in total, $H_0$, $A$, $f(z_1)$, $f(z_2)$, $f(z_3)$, $f(z_4)$, and $f(z_5)$. Note that $f(z)$ acts as a one-point function in contrast to $Om3$ which is a three-point function. 

If we rewrite (\ref{eq:om3a}) in terms of $f(z)$:

\begin{equation}
Om3(z_i;z_j;z_k) =
\dfrac{A + \dfrac{f(z_j)-f(z_i)}{(1+z_j)^3-(1+z_i)^3}}
{A + \dfrac{f(z_k)-f(z_i)}{(1+z_k)^3-(1+z_i)^3}},
\label{eq:(4)}
\end{equation}
where $Om3$ is expressed in terms of $f(z)$ functions. Now we make some model assumptions to understand if we can get meaningful information from these $f(z)$ functions and if they can show deviations from expectations of the $\Lambda$CDM model similar to what $Om3$ performs. We 
assume two fiducial cosmology models, one being the standard $\Lambda$CDM model with $w=-1$ and $\Omega_m=0.3$, and another one being a $w$CDM model with constant $w=-0.905$ and $\Omega_m=0.285$. We also assume values of $H_0$ = 67.37 km/s/Mpc and $r_d$ = 147.18 Mpc for both models~\cite{Planck:2018vyg}. 

Now dealing with these models and since $H_0$ and $A$ are unknown, we assume four arbitrary sets of these parameters and derive $f(z)$ functions and different $Om3$ combinations for each set. Note that the purpose of the analysis in this appendix is to show the sensitivity of $f(z)$ and $Om3$ to the assumptions of cosmological parameters (here $H_0$ and $A$) and see if we can get any insight about the actual cosmological models. Hence we did not consider uncertainties of the data in this part. 


The results for $f(z)$ and $Om3$ are shown in Table~\ref{table:tableapp}. We can see that the derived values of $f(z)$ are very much dependent on the assumed values of $H_0$ and $A$, also in the case of $\Lambda$CDM model, we cannot rely on the $f(z)$ values since having an incorrect assumption of $H_0$ and $A$ result to nonzero values of $f(z)$ functions that can be interpreted as a deviation from $\Lambda$CDM model. Contrary to $f(z)$ functions, $Om3$ [that can also be derived from $f(z)$ values] is completely independent of the choices of $H_0$ and $A$ and can indicate consistency with $\Lambda$CDM model when the model is actually $\Lambda$CDM. $Om3$ can also show deviation from $\Lambda$CDM model when the fiducial model is not $\Lambda$CDM. 

In conclusion, $f(z)$ values vary with the choice of $H_0$ and $A$ combination. However, $Om3$ values are independent of $H_0$ and $A$ by construction. This illustrates the strength of the {\em model independent} $Om3$ diagnostic test against simple one-point parametric (or nonparametric) functions. 


\bibliography{apssamp}

\end{document}